\documentclass{article}

\PassOptionsToPackage{numbers, compress}{natbib}
\usepackage{graphicx}

\usepackage[final]{neurips_2021}

\usepackage[utf8]{inputenc} % allow utf-8 input
\usepackage[T1]{fontenc}    % use 8-bit T1 fonts
\usepackage{hyperref}       % hyperlinks
\usepackage{url}            % simple URL typesetting
\usepackage{booktabs}       % professional-quality tables
\usepackage{amsfonts}       % blackboard math symbols
\usepackage{nicefrac}       % compact symbols for 1/2, etc.
\usepackage{microtype}      % microtypography
\usepackage{xcolor}         % colors
\usepackage{caption}
\usepackage{subcaption}
\usepackage[export]{adjustbox}

\usepackage{todonotes}

\title{Incorporating High-Frequency Weather Data into Consumption Expenditure Predictions}

\author{%
  Anders ~Christensen \\
  Section for Cognitive Systems\\
  Technical University of Denmark\\
  Copenhagen, Denmark \\
  \texttt{andchri@dtu.dk} \\
  \And
  Joel ~Ferguson\thanks{https://joelferg.github.io/}\\
  Department of Agricultural and Resource Economics \\
  University of California, Berkeley \\
  Berkeley, CA 94704\\
  \texttt{joel\_ferg@berkeley.edu} \\
  \And
  Sim\'{o}n ~Ram\'{i}rez Amaya\\
  School of Information \\
  University of California, Berkeley \\
  Berkeley, CA 94704\\
  \texttt{simonra@berkeley.edu} \\
}

\begin{document}

\maketitle

\begin{abstract}
  Recent efforts have been very successful in accurately mapping welfare in data-sparse regions of the world using satellite imagery and other non-traditional data sources. However, the literature to date has focused on predicting a particular class of welfare measures, asset indices, which are relatively insensitive to short-term fluctuations in well-being. We suggest that predicting more volatile welfare measures, such as consumption expenditure, substantially benefits from the incorporation of data sources with high temporal resolution. By incorporating daily weather data into training and prediction, we improve consumption prediction accuracy significantly compared to models that only utilize satellite imagery.
\end{abstract}

\section{Introduction}

A surge of recent research has shown that machine learning applied to satellite imagery can predict socio-economic characteristics at a fine spatial scale with reasonable accuracy \citep{Jean2016,Babenko2017,Perez2017,Steele2017,Han2019,Piaggesi2019,Yeh2020}. In data-scarce parts of the world, these prediction methods have the potential to dramatically increase policymakers' understanding of the spatial dimensions of economic well-being, and support decisions about targeting poverty assistance, program implementation choices, and evaluation. Satellite imagery-based predictions over a wide variety of tasks have been shown to perform reasonably well -- and the state of the science is quickly improving \cite{Burke2021}.

\begin{figure}
  \label{fig:withinvar}
  \centering
  \includegraphics[width = \textwidth]{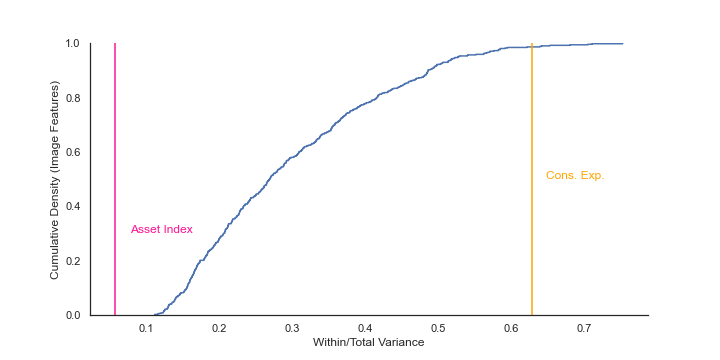}
  \caption{This figure shows the empirical cumulative density function of the ratio of within-sum-of-squares to total-sum-of-squares for the 512 satellite imagery features outputted by the convolutional neural network trained by \citep{Yeh2020}. The pink vertical line shows the ratio for our asset index measure and the orange vertical line shows the ratio for consumption expenditure. All values were calculated using 588 enumeration area-visit observations, four for each visit to the 147 enumeration areas for which we are able to compute the asset index, consumption expenditure, and image features for the maximum-possible four visits.}
\end{figure}

However, previous welfare mapping exercises have proven especially successful in predicting one particular class of measures: asset indices, usually defined as the first principle component of an asset ownership matrix. These outcomes are attractive for their relative availability of ground truth data, but may omit important aspects of well-being. Asset ownership evolves slowly \cite{Caballero1993} and is generally low in developing country contexts \cite{Banerjee2007}, and so may smooth-over large short-term changes in well-being. As such, asset index predictions are likely less amenable to use in downstream tasks such as identifying locations most severely affected by a shock or determining the short-run outcomes of a policy. Previous studies that have sought to predict per capita consumption expenditure, which is more responsive to transient fluctuations in well-being, alongside an asset index have found substantial performance reductions \citep{Jean2016,Yeh2020}. This may be due to the relatively little within-unit variance in satellite images' signal compared to consumption expenditure, as shown in \hyperref[fig:withinvar]{Figure 1}.

Here, we make progress by incorporating data with high temporal resolution and within-unit variance, particularly daily temperature and precipitation. There is a large literature in development economics that uses weather variation to predict variation in economic activity in agriculture-dependent regions \citep{paxson1992using,miguel2004economic,jayachandran2006selling}. As such, it seems natural that weather would be useful in predicting short-term fluctuations in welfare in such regions. We limit our focus to Nigeria, hypothesizing that our findings will transfer to other countries with economies relying heavily on agriculture. Our results suggest that there are potentially large gains in prediction accuracy for consumption expenditure to be had by incorporating weather and other high-temporal resolution data. We note, however, that there are severe limitations to only utilizing weather data as a source of high-temporal resolution variation. In particular, weather will be uncorrelated with any sort of policy intervention and thus unable to help determine the effects of such interventions. Rather, we seek to emphasize the limited temporal variation in satellite imagery and alternative data sources may be more useful for predicting short-term variation in consumption. Improving short-term welfare prediction is of importance for downstream applications such as targeting for humanitarian assistance or creation of policies, as e.g. conducted in Togo to improve targeting in response to the COVID-19 pandemic \citep{aiken2021machine}.

\section{Data Sources}

\subsection{Ground Truth Welfare Data}

We construct log per capita consumption expenditure measures using data from the first four waves (2010-2011, 2012-2013, 2015-2016, and 2018-2019) of the Nigerian General Household Survey-Panel (GHS), Nigeria’s publicly available Living Standards and Measurement Survey \citep{ghs2010,ghs2012,ghs2015,ghs2018}. In contrast to the Demographic and Health Surveys that have been more widely utilized in the welfare mapping literature, the GHS revisits households. The initial sampling frame for the GHS was a selection of 5,000 households, 10 from each of 500 enumeration areas. These households, as well as any households that split off from the original 5,000, were intended to be surveyed in waves 1, 2, and 3. Wave 4 included a ``partial refresh,” drawing 360 new enumeration areas and keeping 159 of the original enumeration areas (1,507 households). The panel structure of the GHS has the advantage of limiting sampling variation due to changing household and location composition. Additionally, each wave of the GHS is composed of a pre-planting and a post-harvest visit, both of which include a consumption module (asset ownership is only collected in post-planting visits), giving us up to eight observations per household. After calculating consumption expenditure for each household and survey visit, we aggregate them to the enumeration area-visit level by simple averaging and finally take the log.

As asset inventories are only collected once per GHS wave, we also use data from the three most recent publicly available Demographic and Health Survey (DHS) waves from Nigeria (2008, 2013, and 2018) \citep{dhs2008,dhs2013,dhs2018}, to give as approximately as many asset index observations as for the other welfare measures. To ensure we use a consistent asset index measure, we pooled all asset inventories across GHS and DHS waves and construct the measure as the first principal component of the matrix of ownership indicators for assets on which both surveys collect data.

In both the GHS and DHS, enumeration areas' centroid coordinates are displaced up to 10 km to protect the privacy of respondents.

\subsection{Input Data}

All input data was accessed via Google Earth Engine. Multispectral images (MS) used are from Landsat 7 and Nightlights images (NL) come from the Defense Meteorological Program (DMSP) and Visible Infrared Imaging Radiometer Suite (VIIRS). All daytime images have a spatial resolution of 30m/pixel and the following bands: RED, GREEN, BLUE, NIR (Near Infrared), SWIR1 (Shortwave Infrared 1), SWIR2 (Shortwave Infrared 2), and TEMP1 (Thermal). Images used from training and inference are each constructed from one year of image observations by taking the median cloud free pixel from the 365 days preceding the end date of the survey visit of outcome observation. Our image exports took the form of 255 × 255 pixel tiles, that were then center-cropped to 224 × 224 pixels, meaning our final tiles cover a square area with a side length of 6.72 km.

Our weather data are from the ERA5 Reanalysis \cite{hersbach2020era5}, which reports hourly estimates of weather variables on a 0.25 arc-degree grid. We use daily aggregates of total precipitation and mean temperature to construct empirical quintiles of these variables for each of the six months prior to a survey's end date.

\section{Models}

% We evaluate 3 classes of models, each of which is fitted with and without weather data and trained to predict both consumption expenditure and asset index. First, we train a random forest (RF) which takes as features empirical ventiles of the pixel radiance values of each band of the Landsat composite centered on the reported enumeration area centroid. When trained with weather, weather features are concatenated to the image statistic features.

We evaluate two primary classes of models, each of which is fitted with and without weather data and trained to predict both consumption expenditure and asset index.

Firstly, we use the pre-trained convolutional neural net (CNN) trained by \citep{Yeh2020} to predict an asset index for which ground truth observations from Nigeria were in a held out sample. Two CNNs are trained on MS and NL separately. As in \citep{Yeh2020}, we extract and concatenate the features of the pen-ultimate layer of the two models on a fixed set of validation observations and train a ridge regression using these features with the shrinkage parameter chosen via cross-validation. Weather variables are incorporated in the ridge regression step by concatenation with the features from the MS and NL CNNs. We use this setup to predict both asset indices and consumption expenditure.

Secondly, we train the CNNs from \citep{Yeh2020} from scratch to instead predict consumption expenditure in an attempt to optimize the extracted features of the CNNs to be more predictive of consumption expenditure. The model is trained directly on MS and NL data from Nigeria, that has been split into a training, validation and test set by sampling randomly. The pre-trained model is also evaluated on the test data of this split. Weather variables are used as described for the pre-trained model. As a sanity check for this setup and the reduction in data, we train the CNNs from scratch on the same data in order to predict asset indices in Nigeria, and confirm that we get similar accuracies as the pre-trained CNNs.

%For reference, we include a ridge regression predicting consumption expenditure trained just pn the weather data on the corresponding dataset.

% ((((Finally, fit the same CNN model of \citep{Yeh2020} end-to-end before repeating the ridge regression training described in the previous paragraph. Again, weather data are incorporated in the ridge regression step.)))

\begin{figure}[!htbp]
  \label{fig:barperformance}
  \centering
  \includegraphics[width = \textwidth]{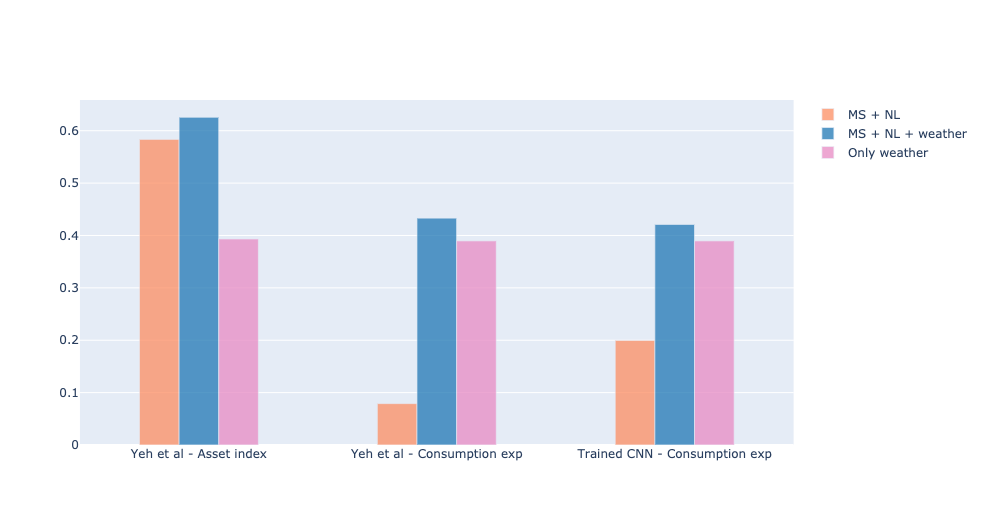}
  \caption{Model performance measured by $R^2$ on the final ridge regression on extracted features from images (MS and NL) and weather. The leftmost bar group presents results for Yeh et al. pretrained CNNs for asset index prediction. The bar group in the center presents results for Yeh et al. pretrained CNNs for consumption expenditure prediction. Finally, the rightmost bar group presents results for our own CNNs trained to predict consumption expenditure. In all cases weather augmented models outperform the other models ($R^2$ values of 0.635,0.433 and 0.421 respectively).}
\end{figure}

\section{Results and discussion}

In \hyperref[fig:barperformance]{Figure 2} we have plotted the performances of the model types described above in terms of $R^2$ on the validation set. Several noteworthy findings seem to appear from this.

\begin{figure}[!htbp]
    \label{fig:predmap}
    \centering
    \begin{subfigure}[b]{1.0\textwidth}
        \adjincludegraphics[scale=0.34,trim={0 {135pt} 0 {120 pt}},clip]{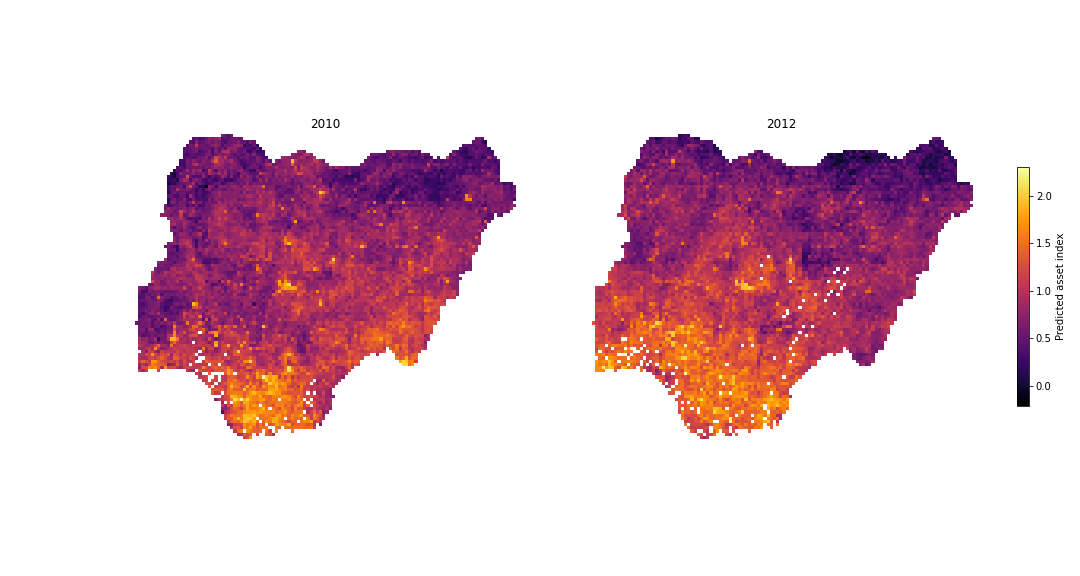}
        \caption{Prediction of asset index}
        \label{rfidtest_xaxis}
    \end{subfigure}
    \begin{subfigure}[b]{1.0\textwidth}
        \adjincludegraphics[scale=0.34,trim={0 {135pt} 0 {80 pt}},clip]{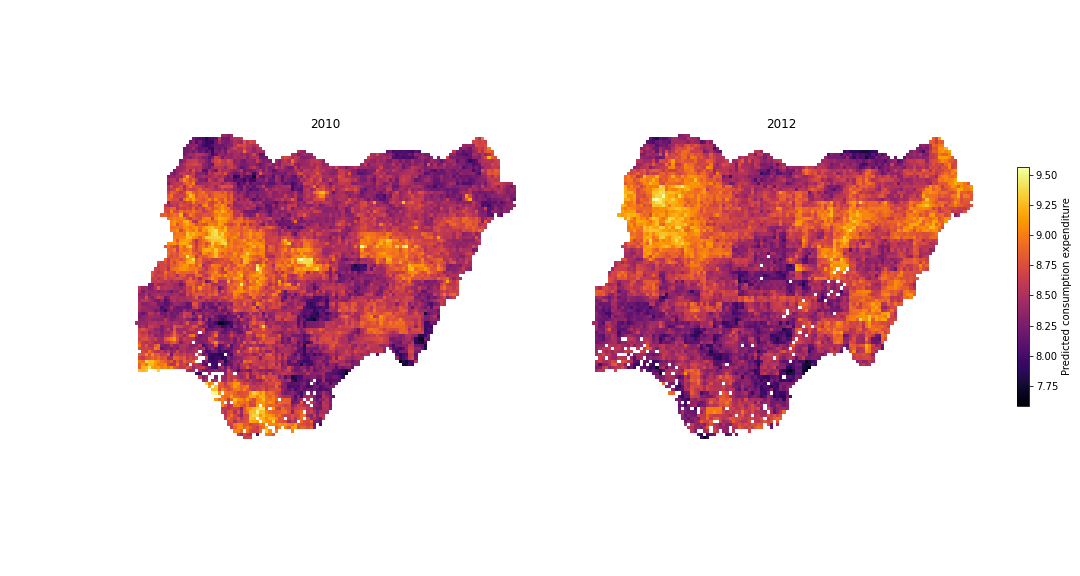}
        \caption{Prediction of consumption expenditure}
        \label{rfidtest_yaxis}
    \end{subfigure}
    \caption{Prediction of asset index (panel a) and log consumption expenditure (panel b) for every 0.1 arc-degree grid cell in Nigeria for 2010 and 2012 using the pretrained CNNs of \citep{Yeh2020} and weather data. Cells for which input data was not available are left blank.}
    \label{rfidtag_testing}
\end{figure}

Firstly, although the features derived from the pretrained CNNs of \citep{Yeh2020} are predictive of asset index, which they are trained upon, the information these CNNs are extracting from the images do not seem to be predictive of consumption expenditure. This indicates that features for predicting asset index are unable to be used for prediction of the more volatile consumption expenditure. However, training the CNNs to generate features predictive of consumption expenditure rather than asset index only slightly increases performance. This warns of the image data modality being lackluster for predicting the volatile consumption expenditure.

For both of these setups, however, we show that simply including high temporal resolution weather data in the final ridge regression on the extracted features boosts performance significantly. On the other hand, performance of the asset index model only increases a small amount with this additional information.

One curious point appears when comparing the pretrained model and the model trained to predict consumption expenditure when using all three data modalities. Here, performance is actually slightly lower for the model trained to extract features from the imagery that is directly predictive of consumption expenditure. We hypothesize that this is due to the asset index CNNs extracting features predictive of long-term wealth, whereas the consumption expenditure CNNs attempt to extract relevant information for a shorter period of time. Combining the extracted features from the images with weather data with higher temporal resolution, we see the best performance when combining long- and short-term information. We hypothesize that training the full model in an end-to-end fashion by incorporating all our discussed data modalities jointly will improve results further.

\section{Conclusion}

Our results suggest that weather data is predictive of consumption expenditure and that weather-augmented models outperform common alternatives in the literature. As shown in \hyperref[fig:predmap]{Figure 3}, this leads to meaningfully different predictions regarding the distribution of welfare in space and how it changes over time. Strong correlations between weather and economic activity have been documented in agrarian societies across the world \citep{damania2020does}. As such, we would expect this pattern of results to hold in many low-resource and data-scarce settings.

As noted above, weather data will likely be unpredictive of other important shocks to consumption (e.g. the COVID-19 pandemic), limiting its effectiveness as a data source. An avenue for moving research forward is to find data sources that, like satellite imagery, are potentially related to many facets of economic well-being, but are available at high temporal resolution and carry substantial within-unit variance. Call detail records, for example, have also been used to predict welfare measures accurately and plausibly exhibit much more within-unit variation \citep{Blumenstock2015}.

We would like to also stress the importance of considering alternative measures of welfare in ML-based welfare mapping exercises. Consumption expenditure does not account for the composition of consumption and incorporates price information that is uninformative of welfare. To illustrate this, consider two households that consume the exact same basket of goods but face different prices. One household would generally appear to have greater welfare as measured by consumption expenditures, despite the households having equal real consumption. Two meaningful direction for future research are (i) explore whether more theoretically-grounded measures of welfare (e.g. marginal utility of expenditures (IMUE) as in \citep{Ligon2019}) can be predicted with similar accuracy and (ii) explore whether the choice of welfare measure leads to systematic differences in downstream applications.

Finally, we would like to mention that while welfare mapping can provide valuable insight into the spatial and temporal distribution of well-being, all welfare measures are imperfect as illustrated by the example above. Furthermore, prediction errors always distort our picture of welfare, regardless of the measure of interest. Blind application of welfare predictions can lead to very misleading policy recommendations or evaluations and practitioners should be aware of the limitations of this approach.

\clearpage
\bibliography{ml4d_submission.bib}

\begin{thebibliography}{25}
\providecommand{\natexlab}[1]{#1}
\providecommand{\url}[1]{\texttt{#1}}
\expandafter\ifx\csname urlstyle\endcsname\relax
  \providecommand{\doi}[1]{doi: #1}\else
  \providecommand{\doi}{doi: \begingroup \urlstyle{rm}\Url}\fi

\bibitem[Jean et~al.(2016)Jean, Burke, Xie, Davis, Lobell, and Ermon]{Jean2016}
Neal Jean, Marshall Burke, Michael Xie, W.~Matthew Davis, David~B. Lobell, and
  Stefano Ermon.
\newblock {Combining satellite imagery and machine learning to predict
  poverty}.
\newblock \emph{Science}, 353\penalty0 (6301):\penalty0 790--794, 2016.
\newblock ISSN 10959203.
\newblock \doi{10.1126/science.aaf7894}.

\bibitem[Babenko et~al.(2017)Babenko, Hersh, Newhouse, Ramakrishnan, and
  Swartz]{Babenko2017}
Boris Babenko, Jonathan Hersh, David Newhouse, Anusha Ramakrishnan, and Tom
  Swartz.
\newblock {Poverty mapping using convolutional neural networks trained on high
  and medium resolution satellite images, with an application in Mexico}.
\newblock \emph{arXiv}, \penalty0 (Nips):\penalty0 1--4, 2017.
\newblock ISSN 23318422.

\bibitem[Perez et~al.(2017)Perez, Yeh, Azzari, Burke, Lobell, and
  Ermon]{Perez2017}
Anthony Perez, Christopher Yeh, George Azzari, Marshall Burke, David Lobell,
  and Stefano Ermon.
\newblock {Poverty prediction with public landsat 7 satellite imagery and
  machine learning}.
\newblock \emph{arXiv}, \penalty0 (Nips), 2017.
\newblock ISSN 23318422.

\bibitem[Steele et~al.(2017)Steele, Sunds{\o}y, Pezzulo, Alegana, Bird,
  Blumenstock, Bjelland, Eng{\o}-Monsen, de~Montjoye, Iqbal, Hadiuzzaman, Lu,
  Wetter, Tatem, and Bengtsson]{Steele2017}
Jessica~E. Steele, P{\aa}l~Roe Sunds{\o}y, Carla Pezzulo, Victor~A. Alegana,
  Tomas~J. Bird, Joshua Blumenstock, Johannes Bjelland, Kenth Eng{\o}-Monsen,
  Yves-Alexandre de~Montjoye, Asif~M. Iqbal, Khandakar~N. Hadiuzzaman, Xin Lu,
  Erik Wetter, Andrew~J. Tatem, and Linus Bengtsson.
\newblock {Mapping poverty using mobile phone and satellite data}.
\newblock \emph{Journal of The Royal Society Interface}, 14\penalty0
  (127):\penalty0 20160690, feb 2017.
\newblock ISSN 1742-5689.
\newblock \doi{10.1098/rsif.2016.0690}.
\newblock URL
  \url{https://royalsocietypublishing.org/doi/10.1098/rsif.2016.0690}.

\bibitem[Han et~al.(2019)Han, Ahn, Cha, Yang, Park, and Cha]{Han2019}
Sungwon Han, Donghyun Ahn, Hyunji Cha, Jeasurk Yang, Sungwon Park, and Meeyoung
  Cha.
\newblock {Lightweight and robust representation of economic scales from
  satellite imagery}.
\newblock \emph{arXiv}, 2019.
\newblock ISSN 23318422.
\newblock \doi{10.1609/aaai.v34i01.5379}.

\bibitem[Piaggesi et~al.(2019)Piaggesi, Gauvin, Tizzoni, Adler, Verhulst,
  Young, Price, Ferres, Cattuto, and Panisson]{Piaggesi2019}
Simone Piaggesi, Laetitia Gauvin, Michele Tizzoni, Natalia Adler, Stefaan
  Verhulst, Andrew Young, Rihannan Price, Leo Ferres, Ciro Cattuto, and
  Andr{\'{e}} Panisson.
\newblock {Predicting City Poverty Using Satellite Imagery}.
\newblock pages 90--96, 2019.
\newblock URL \url{https://censusreporter.org/topics/income/}.

\bibitem[Yeh et~al.(2020)Yeh, Perez, Driscoll, Azzari, Tang, Lobell, Ermon, and
  Burke]{Yeh2020}
Christopher Yeh, Anthony Perez, Anne Driscoll, George Azzari, Zhongyi Tang,
  David Lobell, Stefano Ermon, and Marshall Burke.
\newblock {Using publicly available satellite imagery and deep learning to
  understand economic well-being in Africa}.
\newblock \emph{Nature Communications}, 11\penalty0 (1):\penalty0 1--11, 2020.
\newblock ISSN 20411723.
\newblock \doi{10.1038/s41467-020-16185-w}.
\newblock URL \url{http://dx.doi.org/10.1038/s41467-020-16185-w}.

\bibitem[Burke et~al.(2021)Burke, Driscoll, Lobell, and Ermon]{Burke2021}
Marshall Burke, Anne Driscoll, David~B. Lobell, and Stefano Ermon.
\newblock {Using satellite imagery to understand and promote sustainable
  development}.
\newblock \emph{Science}, 371\penalty0 (6535), 2021.
\newblock ISSN 10959203.
\newblock \doi{10.1126/science.abe8628}.

\bibitem[Caballero(1993)]{Caballero1993}
Ricardo~J Caballero.
\newblock {Durable Goods: An Explanation for Their Slow Adjustment}.
\newblock \emph{Journal of Political Economy}, 101\penalty0 (2):\penalty0
  351--384, 1993.
\newblock ISSN 0022-3808.
\newblock \doi{10.1086/261879}.

\bibitem[Banerjee and Duflo(2007)]{Banerjee2007}
Abhijit~V. Banerjee and Esther Duflo.
\newblock {The economic lives of the poor}.
\newblock \emph{Journal of Economic Perspectives}, 21\penalty0 (1):\penalty0
  141--167, 2007.
\newblock ISSN 08953309.
\newblock \doi{10.1257/jep.21.1.141}.

\bibitem[Paxson(1992)]{paxson1992using}
Christina~H Paxson.
\newblock Using weather variability to estimate the response of savings to
  transitory income in thailand.
\newblock \emph{The American Economic Review}, pages 15--33, 1992.

\bibitem[Miguel et~al.(2004)Miguel, Satyanath, and
  Sergenti]{miguel2004economic}
Edward Miguel, Shanker Satyanath, and Ernest Sergenti.
\newblock Economic shocks and civil conflict: An instrumental variables
  approach.
\newblock \emph{Journal of Political Economy}, 112\penalty0 (4):\penalty0
  725--753, 2004.

\bibitem[Jayachandran(2006)]{jayachandran2006selling}
Seema Jayachandran.
\newblock Selling labor low: Wage responses to productivity shocks in
  developing countries.
\newblock \emph{Journal of Political Economy}, 114\penalty0 (3):\penalty0
  538--575, 2006.

\bibitem[Aiken et~al.(2021)Aiken, Bellue, Karlan, Udry, and
  Blumenstock]{aiken2021machine}
Emily Aiken, Suzanne Bellue, Dean Karlan, Christopher~R Udry, and Joshua
  Blumenstock.
\newblock Machine learning and mobile phone data can improve the targeting of
  humanitarian assistance.
\newblock Technical report, National Bureau of Economic Research, 2021.

\bibitem[of~Statistics(2010)]{ghs2010}
Nigeria National~Bureau of~Statistics.
\newblock General household survey, panel (ghs-panel) 2010-2011, 2010.
\newblock Dataset downloaded from www.microdata.worldbank.org on January 11,
  2021.

\bibitem[of~Statistics(2012)]{ghs2012}
Nigeria National~Bureau of~Statistics.
\newblock General household survey, panel (ghs-panel) 2012-2013, 2012.
\newblock Dataset downloaded from www.microdata.worldbank.org on January 24,
  2021.

\bibitem[of~Statistics(2015)]{ghs2015}
Nigeria National~Bureau of~Statistics.
\newblock General household survey, panel (ghs-panel) 2015-2016, 2015.
\newblock Dataset downloaded from www.microdata.worldbank.org on January 24,
  2021.

\bibitem[of~Statistics(2018)]{ghs2018}
Nigeria National~Bureau of~Statistics.
\newblock General household survey, panel (ghs-panel) 2018-2019, 2018.
\newblock Dataset downloaded from www.microdata.worldbank.org on January 11,
  2021.

\bibitem[NPC/Nigeria and Macro(2009)]{dhs2008}
National Population~Commission NPC/Nigeria and ICF Macro.
\newblock Nigeria demographic and health survey 2008 [dataset], 2009.
\newblock Abuja, Nigeria: NPC/Nigeria and ICF Macro [Producers]. IFC
  [Distributor],2009.

\bibitem[NPC/Nigeria and International(2014)]{dhs2013}
National Population~Commission NPC/Nigeria and ICF International.
\newblock Nigeria demographic and health survey 2013 [dataset], 2014.
\newblock AAbuja, Nigeria: NPC/Nigeria and ICF International [Producers]. IFC
  [Distributor],2014.

\bibitem[NPC/Nigeria and ICF(2019)]{dhs2018}
National Population~Commission NPC/Nigeria and ICF.
\newblock Nigeria demographic and health survey 2018 [dataset], 2019.
\newblock Abuja, Nigeria, and Rockville, Maryland, USA: NPC and ICF
  [Producers]. IFC [Distributor],2019.

\bibitem[Hersbach et~al.(2020)Hersbach, Bell, Berrisford, Hirahara,
  Hor{\'a}nyi, Mu{\~n}oz-Sabater, Nicolas, Peubey, Radu, Schepers,
  et~al.]{hersbach2020era5}
Hans Hersbach, Bill Bell, Paul Berrisford, Shoji Hirahara, Andr{\'a}s
  Hor{\'a}nyi, Joaqu{\'\i}n Mu{\~n}oz-Sabater, Julien Nicolas, Carole Peubey,
  Raluca Radu, Dinand Schepers, et~al.
\newblock The era5 global reanalysis.
\newblock \emph{Quarterly Journal of the Royal Meteorological Society},
  146\penalty0 (730):\penalty0 1999--2049, 2020.

\bibitem[Damania et~al.(2020)Damania, Desbureaux, and Zaveri]{damania2020does}
Richard Damania, Sebastien Desbureaux, and Esha Zaveri.
\newblock Does rainfall matter for economic growth? evidence from global
  sub-national data (1990--2014).
\newblock \emph{Journal of Environmental Economics and Management},
  102:\penalty0 102335, 2020.

\bibitem[Blumenstock et~al.(2015)Blumenstock, Cadamuro, and
  On]{Blumenstock2015}
Joshua Blumenstock, Gabriel Cadamuro, and Robert On.
\newblock {Predicting poverty and wealth from mobile phone metadata}.
\newblock \emph{Science}, 350\penalty0 (6264):\penalty0 1073--1076, 2015.
\newblock ISSN 10959203.
\newblock \doi{10.1126/science.aac4420}.

\bibitem[Ligon(2019)]{Ligon2019}
Ethan Ligon.
\newblock {Estimating Household Welfare from Disaggregate Expenditures}.
\newblock \emph{Working Paper}, 2019.
\newblock ISSN 1884-5088.

\end{thebibliography}

\end{document}